\documentclass[draftcls, journal, onecolumn,12pt]{IEEEtran}
\usepackage{cite}

\ifCLASSINFOpdf
   \usepackage[pdftex]{graphicx}
   \graphicspath{{../pdf/}{../jpeg/}}
   \DeclareGraphicsExtensions{.pdf,.jpeg,.png}
\else
   \usepackage[dvips]{graphicx}
   \graphicspath{{../eps/}}
   \DeclareGraphicsExtensions{.eps}
\fi
\usepackage[cmex10]{amsmath}

\usepackage{algorithm}
\usepackage{algorithmic}
\begin{document}
%
\title{QoS-Aware User Association for Load Balancing in Heterogeneous Cellular Networks}
%
%
%

\author{Tianqing Zhou,
        Yongming Huang,~\IEEEmembership{Member,~IEEE}
        and Luxi Yang,~\IEEEmembership{Member,~IEEE,}
\thanks{T. Zhou, Y. Huang and L. Yang are with the School of Information Science
and Engineering, Southeast University, Nanjing 210096, China. They are
also with the Key Laboratory of Underwater Acoustic Signal Processing
of Ministry of Education, Southeast University (e-mail: {zhoutian930}@163.com, {\{huangym, lxyang\}}@seu.edu.cn.}}
\maketitle

\begin{abstract}
To solve the problem that the low capacity in hot-spots and coverage holes of conventional cellular networks, the base stations (BSs) having lower transmit power are deployed to form heterogeneous cellular networks (HetNets). However, because of these introduced disparate power BSs, the user distributions among them looked fairly unbalanced if an appropriate user association scheme hasn't been provided. For effectively tackling this problem, we jointly consider the load of each BS and user's achievable rate instead of only utilizing the latter when designing an association algorithm, and formulate it as a network-wide weighted utility maximization problem.  Note that, the load mentioned above relates to the amount of required subbands decided by actual rate requirements, i.e., QoS,  but the number of associated users, thus it can reflect user's actual load level. As for the proposed problem, we give a maximum probability (max-probability) algorithm by relaxing variables as well as a low-complexity distributed algorithm with a near-optimal solution that provides a theoretical performance guarantee. Experimental results show that, compared with the association strategy advocated by Ye, our strategy has a speeder convergence rate, a lower call blocking probability and a higher load balancing level.
\end{abstract}


%
\IEEEpeerreviewmaketitle

\section{Introduction}
%
%
%
%
\IEEEPARstart{T}{o} keep pace with explosive growth in data traffic demands driven by various wireless user equipments (especially media-hungry devices), network operators have to take into account increasing the network capacity and reducing the cost/bit delivered by perhaps two orders of magnitude \cite{1}. According to the visual networking index revealed by Cisco \cite{2}, global mobile data traffic grew 2.6-fold in 2010 and would increase by up to 26 times in the next five years. Obviously, the typical cautious schemes to increase network capacity can't respond magnificently to this challenge.
\par
Since original voice-oriented wireless services change into being data-oriented and more user equipments operate indoors, more link budget and coverage extension are needed to meet more user requirements \cite{3}. To do this cost-effectively, the infrastructure deployment of a cellular network is trending strongly towards having heterogeneous elements and away from conventional high-power base stations (BSs). These elements mainly include micro, pico, femto and relay BSs, which differ primarily in transmit power, physical size, backhaul, cost, ease-of-deployment and propagation characteristics \cite{4,5,6}. As is well known, the rational deployment of low-power BSs would be beneficial to eliminate coverage holes in the conventional cellular network and improve capacity in hot-spots \cite{7} However, as the signal strength user received from different BSs is heavily dependent on transmit power of BSs, the coverage region of small BSs will be smaller than higher power BSs if we only adopt the maximum signal strength association strategies.  For making full use of new low-power infrastructures, we should consider a proper user association algorithm that actively pushes mobile users onto the lightly loaded small BS, which brings higher rates over time by supplying mobile users with more physic resources. Clearly, a balanced user association scheme should be able to reduce the load of high-power infrastructures, providing more satisfactory services for its remaining users.
\subsection{Related Work}
As the heterogeneity of BSs deployed in heterogeneous cellular networks (HetNets), the association scheme that only depend on user's received signal strength, such as maximum signal interference noise ratio (SINR), maximum achievable rate, best channel quality and nearest distance, may be no longer appropriate. If these algorithms are adopted when we carry out user association in HetNets, there will be very uneven loads distributed among different BSs even if mobile users are distributed uniformly in geography. That means the cell association in HetNets, compared with conventional cellular networks, becomes more complex and can provide larger potential gains by utilizing load-aware associations.
\par
As far as the load is concerned, there are two types of definitions, which include considering the amount of resources consumed by each BS as the load and adopting the number of users associated with each BS as the load. Therefore, the existing work on user association for load balancing can be broadly classified into two groups:
\par
\begin{enumerate}
\item Strategies based on the number of consumed resources, such as association control \cite{8},   distributed $\alpha$-optimal user association \cite{9}, cell range assignment using cell-specific offset \cite{10}, dynamic load balancing of integrated wireless networks \cite{11}, coordinated scheduling across a cluster of BSs \cite{12}, cell breathing \cite{13,14}, etc;
\item Strategies based on the number of associated users, such as network-wide utility maximization with interference avoidance \cite{15}, aggregate utility of overall rate maximization \cite{16} and biasing methods \cite{17,18,19,20}.
\end{enumerate}
\par
Clearly, the former takes user association according to users' practical requirements and can achieve a genuine load balancing. However, the latter only regards the number of associated users as the load, which may result in a high call blocking rate and unbalanced user distributions among different BSs because of the insufficiency of resources. The proposed approach in this paper is based on the first case, where the frequency band is seen as the load.
\par
In emerging wireless networks, Due to the disparate transmit powers and station capabilities, some load balancing techniques, which are well applied to conventional networks, may become exceedingly improper. As a hot topic in HetNets, the user association for load balancing has so far been far from being full understood \cite{21}. Recently, there have many efforts in the literature toward user distribution balancing. The so-called biasing method \cite{17,18,19,20}, which is most frequently utilized, adds an offset or bias to lower-power BS so that more users can be associated with the lower-power BS. This approach is simple and effective, but it is often inclined to adopting the number of associated users as the load and provides very few ideal ways to obtain an optimal offset or bias factor. Ye et al. \cite{16} proposed a distributed user association algorithm to balance the load defined by the number of users among different BSs. As regards the first strategy, there are few studies for HetNets. Siomina et al. \cite{10} put forward a heuristic load balancing algorithm which can be used for range optimization in HetNets, but it appears to occupy high computational overhead when the number of BSs involved in a cellular system is large. Therefore, how to design a high effective association algorithm for HetNets is still an important open problem.
\subsection{Contributions and Organization}
In this paper, we present a load-aware user association scheme that regards the frequency band as the load  under the restriction of resources for downlink HetNets, which brings the following main contributions.
\par
First, like most of the solving processes, we convert the original problem into a convex optimization one by relaxing association indicator variables and associate users to some BS having a maximal association indicator taken from the solutions of the convex optimization problem. These relaxed association indicator variables are seen as the user association probabilities by us and thus this association scheme is called as the maximum probability (max-probability) association.
\par
Second, motivated by Ye's work \cite{16}, we also introduce a Lagrange multiplier to relax the coupled constraint involved in above convex optimization, and then obtain a distributed association scheme which can provide a near-optimal solution with a theoretical performance guarantee and be well applied to HetNets due to its feasibility, efficiency and low overhead.
\par
Third, we compare the proposed distributed association, max-probability association  maximum achievable rate (max-rate) association and Ye's association algorithm, and then reveal the relationships among them.
\par
The remainder of this paper is organized as follows. In section II, we present our system model. In section III, we formulate the user association problem based on users and on resources. In section IV, we give a max-probability and a distributed association scheme for solving the problem based on resources. In section V, we discuss our simulations in terms of the call blocking probability, load balancing level and convergence for two-tier HetNets. In section VI, we present further discussion and conclusions.
\section{SYSTEM MODEL}
In conventional cellular networks, users are often associated with some BS which can provide the maximum SINR. Obviously, the association scheme indeed maximizes the SINR coverage probability for a downlink system, and typically is the same with nearest neighbor association since a BS with the strongest signal is also the closest one. Having almost the same coverage area for all BSs, this association strategy can be efficacious. However, it doesn't work well for HetNets because the downlink coverage area of BSs with disparate transmit powers is vastly different. Therefore, a key metric for performance ( i.e., the average long-term throughput) should be considered when we design a user association algorithm, which relates to user's achievable rate and actual load of BS.
\par
In this paper, we consider a downlink cell association for HetNets. For the subsequent work, we shall make the following assumption.
\par
\noindent
\itshape \textbf{Assumption:}
\upshape
Power is equally allocated to all employed subbands for each BS.
\par
\noindent
Remark: with implementation simplicity and analytical tractability, this assumption  has been widely used in downlink resource allocation problems \cite{22,23}. Moreover, several studies show that a near-optimal solution can be obtained by using equal power allocation in many cases  especially in high SINR regime \cite{24,25}.
\par
\begin{figure}[!t]
\centering
\centerline{\includegraphics{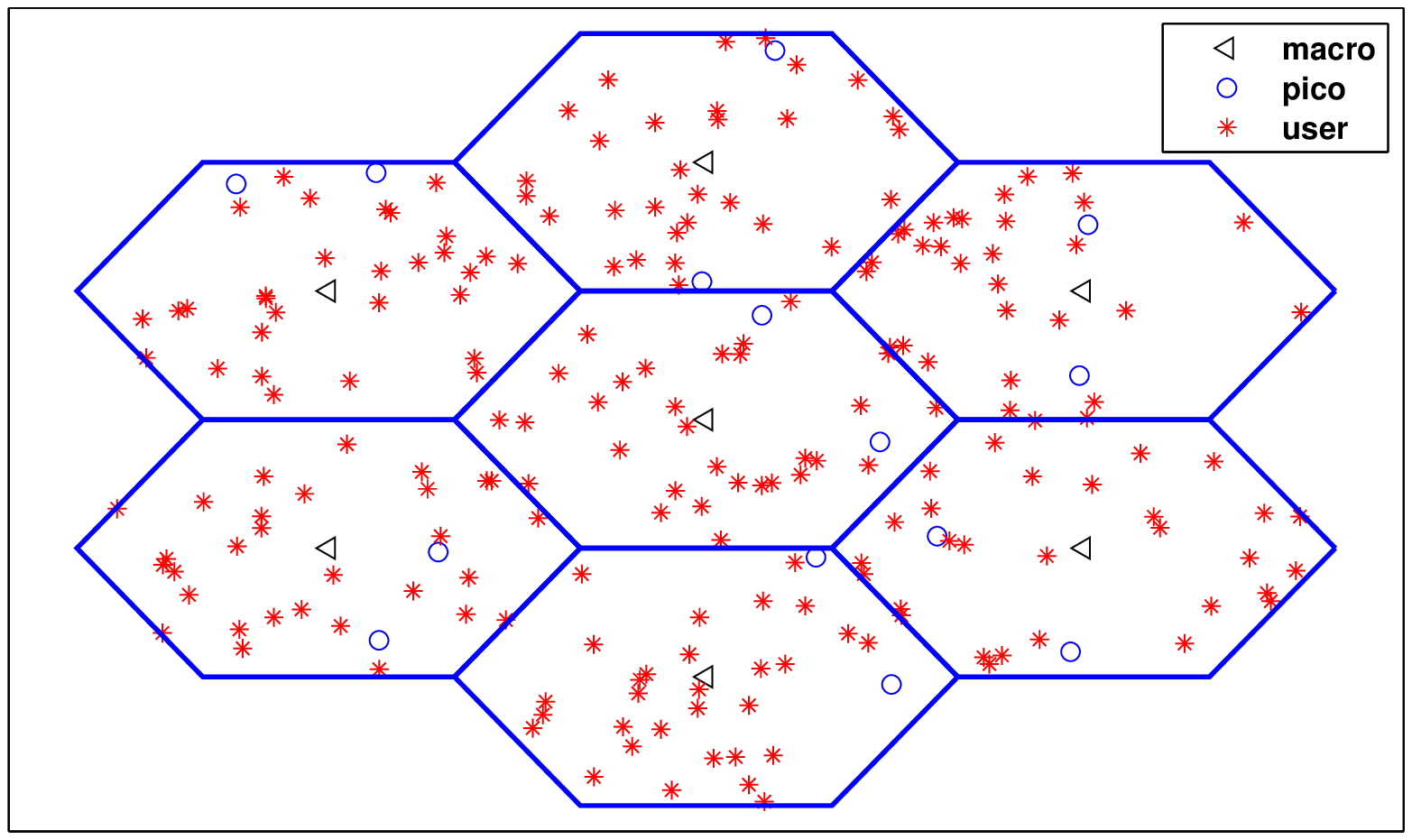}}
\caption{There is a realization of HetNets: macro BSs are regularly distributed to form a cellular system, users and pico BSs are scattered into these macrocells in a random way.}
\label{fig1}
\end{figure}

A  HetNets \cite{26} consisting of various BSs is illustrated in Fig.\ref{fig1}, where macrocells form a regular cellular network, small cells and users are randomly distributed in each macrocell. Without loss of generality, we only refer to one kind of small cells (i.e., picocells) which own smaller coverage than macrocells.
\par
We denote the set of users and the set of BSs including macro BSs and pico BSs in the network by $\mathcal{K}$ and $\mathcal{N}$, respectively. The received SINR at time slot $t$ for user $k\in\mathcal{K}$ from BS $n\in\mathcal{N}$ on one subband including twelve carriers can be written as:
\begin{equation}\label{eq1}
\begin{IEEEeqnarraybox}[][c]{l}
{SINR}_{nk}\left( t \right)=\frac{{{p}_{n}}{{g}_{nk}}\left( t \right)}{\sum\limits_{j\in \mathcal{N}\backslash \left\{ n \right\}}{{{p}_{j}}{{g}_{jk}}\left( t \right)+{{\sigma }^{2}}}}
\end{IEEEeqnarraybox}
\end{equation}
where ${{p}_{n}}{{g}_{nk}}\left( t \right)$ is the received signal strength of user $k$ from BS $n$ at time slot $t$ with ${{p}_{n}}$ and ${{g}_{nk}}$ representing the nonnegative transmit power on one subband of BS $n$ and the channel gain between BS $n$ and user $k$, respectively; ${\sigma }^{2}$ denotes the noise power of each subband.
\par
For a given ${{p}_{n}}$, we can obtain the instantaneous achievable rate [in Kbps] at time slot $t$ for user $k$ from BS $n$ according to the following formula:
\begin{equation}\label{eq2}
\begin{IEEEeqnarraybox}[][c]{l}
{{r}_{nk}}\left( t \right)=W\log 2\left( 1+SIN{{R}_{nk}}\left( t \right) \right)
\end{IEEEeqnarraybox}
\end{equation}
where $W$ denotes the width of subband (180KHz).
\par
For the subsequent study, we should introduce the following theorem from Son's \cite{15} and Kushner's work \cite{27}, where its proof is omitted since they have provided detailed explanations.
\newtheorem{Theorem}{\bfseries \upshape Theorem}
\begin{Theorem}
{\itshape Under some assumptions, if the proportional fair algorithm is adopted as an intra-cell scheduler, the average long-term throughput of user $k$ as $t\to \infty $ can be written as follows:}
\end{Theorem}
\begin{equation}\label{eq3}
\begin{IEEEeqnarraybox}[][c]{l}
{{T}_{k}}=\sum\limits_{n\in \mathcal{N}}{{{x}_{nk}}\frac{J\left( {{y}_{n}} \right){{{\bar{r}}}_{nk}}}{{{y}_{n}}}}
\end{IEEEeqnarraybox}
\end{equation}
\itshape where ${{x}_{nk}}$ denotes the association indicator, i.e., ${{x}_{nk}}=1$ when user $k$ is associated with BS $n$, 0 otherwise; ${{y}_{n}}=\sum\nolimits_{k\in \mathcal{K}}{{{x}_{nk}}}$ represents the number of users associated with BS $n$ ; $J\left( {y}_{n} \right)=\sum\nolimits_{k=1}^{{y}_{n}}{\frac{1}{k}}$ represents a multi-user diversity gain only decided by the number of users competing for the same resource; ${{\bar{r}}_{nk}}$ is the expectation of ${{r}_{nk}}$, i.e., a long-term average about instantaneous achievable data rate of user $k$ on BS $n$.
\upshape
\section{PROBLEM FORMULATION}
In this paper, we focus on an optimization problem of the cell association, which maximizes the network-wide utility. Before formulating the optimization problem, we need to give a following definition:
\newtheorem{Definition}{\bfseries \upshape Definition}
\begin{Definition}
{\itshape If user $k$ is associated with BS $n$, whose average long-term throughput from BS $n$ is}
\end{Definition}
\begin{equation}\label{eq4}
\begin{IEEEeqnarraybox}[][c]{l}
{{c}_{nk}}=\frac{J\left( {{y}_{n}} \right)f\left( {{{\bar{r}}}_{nk}} \right)}{{{y}_{n}}}
\end{IEEEeqnarraybox}
\end{equation}
\itshape
where the function $f$ refers to linear operations of the average long-term throughput to meet requirements of subsequent algorithms, e.g., adding a small enough value to avoid the case $\log \left( {{{\bar{r}}}_{nk}} \right)=-\inf $.
\upshape
\subsection{Optimization Problem based on Users}
Now we formulate an optimization problem that association indicators can be found by maximizing the network-wide aggregate utility function:
\begin{equation}\label{eq5}
\begin{IEEEeqnarraybox}[][c]{cl}
\underset{\mathbf{x}}{\mathop{\max }}&\,\sum\limits_{n\in \mathcal{N}}{\sum\limits_{k\in \mathcal{K}}{{{x}_{nk}}{{U}_{nk}}\left( {{c}_{nk}} \right)}} \\
  \text{    s}\text{.t}\text{.  }&\sum\limits_{k\in \mathcal{K}}{{{x}_{nk}}}={{y}_{n}},\ \ \forall n\in \mathcal{N} \\
 &\sum\limits_{n\in \mathcal{N}}{{{x}_{nk}}}=1,\ \ \forall k\in \mathcal{K} \\
 &{{x}_{nk}}\in \left\{ 0,1 \right\},\ \ \forall n\in \mathcal{N},\forall k\in \mathcal{K}
\end{IEEEeqnarraybox}
\end{equation}
where $\mathbf{x}=\left\{ {{x}_{nk}},n\in \mathcal{N},k\in \mathcal{K} \right\}$ is an indicator vector; ${{y}_{n}}$ is the number of users associated with BS $n$, i.e., load; ${{U}_{nk}}\left( \cdot  \right)$ is a received utility function of user $k$ from BS $n$, which is monotonically increasing, strictly concave, and continuously differentiable; the second constraint $\sum\nolimits_{n\in \mathcal{N}}{{{x}_{nk}}=1}$ shows that each user can be associated with only BS. In this paper, we uniformly adopt a logarithmic function as the utility function mentioned above.
\par
Obviously, the above optimization problem is a familiar 0-1 knapsack and NP-hard problem. In general, achieving its optimal solution is very difficult in reality, especially for the large number of users involved in the network.
\par
We relax association variables and introduce a logarithmic utility function, thus the problem \eqref{eq5} can be rewritten as follows:
\begin{equation}\label{eq6}
\begin{IEEEeqnarraybox}[][c]{cl}
   \underset{\mathbf{x},\mathbf{y}}{\mathop{\max }}&\,\sum\limits_{n\in \mathcal{N}}{\sum\limits_{k\in \mathcal{K}}{{{x}_{nk}}\left\{ \log \left( J\left( {{y}_{n}} \right) \right)+\log \left( {{{\bar{r}}}_{nk}} \right)-\log \left( {{y}_{n}} \right) \right\}}} \\
 \text{    s}\text{.t}\text{.  }&\sum\limits_{k\in \mathcal{K}}{{{x}_{nk}}}={{y}_{n}},\ \ \forall n\in \mathcal{N} \\
 &\sum\limits_{n\in \mathcal{N}}{{{x}_{nk}}}=1,\ \ \forall k\in \mathcal{K} \\
 & 0\le {{x}_{nk}}\le 1,\ \ \forall n\in \mathcal{N},\forall k\in \mathcal{K}
\end{IEEEeqnarraybox}
\end{equation}
where $\mathbf{y}=\left\{ {{y}_{n}},n\in \mathcal{N} \right\}$ is a load vector
\par
Considering the high complexity of above objective function and the practical requirement of implement simplicity, we have to relax objective function to design proper schemes. By the fact $0\le \log \left( J\left( {y}_{n} \right) \right)\le \log \left( {y}_{n} \right)$, we can obtain a maximum rate (max-rate) association problem when $\log\left( J\left( {y}_{n} \right) \right)=\log \left( {y}_{n} \right)$ and the following optimization problem when $\log\left( J\left( {y}_{n} \right) \right)=0$.
\begin{equation}\label{eq7}
\begin{IEEEeqnarraybox}[][c]{cl}
  \underset{\mathbf{x},\mathbf{y}}{\mathop{\max }}&\,\sum\limits_{n\in \mathcal{N}}{\sum\limits_{k\in \mathcal{K}}{{{x}_{nk}}\left\{ \log \left( {{R}_{nk}} \right)-\log \left( {{y}_{n}} \right) \right\}}} \\
 \text{    s}\text{.t}\text{.  }&\sum\limits_{k\in \mathcal{K}}{{{x}_{nk}}}={{y}_{n}},\ \ \forall n\in \mathcal{N} \\
 & \sum\limits_{n\in \mathcal{N}}{{{x}_{nk}}}=1,\ \ \forall k\in \mathcal{K} \\
 & 0\le {{x}_{nk}}\le 1,\ \ \forall n\in \mathcal{N},\forall k\in \mathcal{K}
\end{IEEEeqnarraybox}
\end{equation}
where ${{R}_{nk}}=f\left( {{{\bar{r}}}_{nk}} \right)$. Note that the problem \eqref{eq7} is in accord with the optimization problem proposed by Ye \cite{16} and convex.
\subsection{Optimization Problem based on Resources}
Optimization problem mentioned above defines the number of associated users as the load of each BS, which may results in a high call blocking probability and a low load balancing level. Therefore, it is necessary to design a kind of practical scheme through using the number of consumed resources as the load. Motivated by the problem \eqref{eq7}, we will advocate another more effective strategy instead of it.
\par
Before formulating the optimization problem, we need to calculate the number of consumed resources for each BS according to the practical rate requirement of each user, which is given by
\begin{equation}\label{eq8}
\begin{IEEEeqnarraybox}[][c]{l}
{{s}_{nk}}=\frac{{{d}_{k}}}{{{R}_{nk}}}
\end{IEEEeqnarraybox}
\end{equation}
where ${{d}_{k}}$ is the practical rate of user $k$; ${{s}_{nk}}$ represents the number of required resources.
\par
After calculating the number of required resources, we can give the following definition:
\begin{Definition}
{\itshape If user $k$ is associated with BS $n$, whose load efficiency is}
\end{Definition}
\begin{equation}\label{eq9}
\begin{IEEEeqnarraybox}[][c]{l}
{{e}_{nk}}=\frac{{{R}_{nk}}}{\sum\limits_{j\in \mathcal{K}}{{{x}_{nj}}{{s}_{nj}}}}
\end{IEEEeqnarraybox}
\end{equation}
\par
Now we design a user association scheme by jointly considering achievable rates and practical consumed resources. Therefore, we formulate an optimization problem that association indicators can be found by maximizing the network-wide aggregate weighted utility function:
\begin{equation}\label{eq10}
\begin{IEEEeqnarraybox}[][c]{cl}
  \underset{\mathbf{x}}{\mathop{ \max }}&\,\sum\limits_{n\in \mathcal{N}}{\sum\limits_{k\in \mathcal{K}}{{{x}_{nk}}{{s}_{nk}}{{U}_{nk}}}}\left( {{e}_{nk}} \right) \\
  \text{    s}\text{.t}\text{.  }&\sum\limits_{k\in \mathcal{K}}{{{x}_{nk}}{{s}_{nk}}\le M},\ \ \forall n\in \mathcal{N} \\
 & \sum\limits_{n\in \mathcal{N}}{{{x}_{nk}}}=1,\ \ \forall k\in \mathcal{K} \\
 & {{x}_{nk}}\in \left\{ 0,1 \right\},\ \ \forall n\in \mathcal{N},\forall k\in \mathcal{K}
\end{IEEEeqnarraybox}
\end{equation}
where $\sum\nolimits_{k\in \mathcal{K}}{{{x}_{nk}}{{s}_{nk}}}$ is the number of consumed resources on BS $n$, i.e., the load that is less than the amount of total available resources $M$ (i.e., 100 subbands). Note that the above objective function can be superficially seen as maximizing network-wide weighted load efficiency, where the weight is the corresponding amount of consumed resources.
\par
By relaxing association variables and introducing a logarithmic utility function, the problem \eqref{eq10} can be converted into the following convex optimization problem.
\begin{IEEEeqnarray}{ll}
   \underset{\mathbf{x}}{\mathop{ \max }}&\,\sum\limits_{n\in \mathcal{N}}{\sum\limits_{k\in \mathcal{K}}{{{x}_{nk}}{{s}_{nk}}\left\{ \log \left( {{R}_{nk}} \right)-\log \left( \sum\limits_{j\in \mathcal{K}}{{{x}_{nj}}{{s}_{nj}}} \right) \right\}}} \nonumber \\
 \text{    s}\text{.t}\text{.  }&\sum\limits_{k\in \mathcal{K}}{{{x}_{nk}}{{s}_{nk}}\le M},\ \ \forall n\in \mathcal{N} \nonumber \\
 & \sum\limits_{n\in \mathcal{N}}{{{x}_{nk}}}=1,\ \ \forall k\in \mathcal{K} \label{eq11} \\
 & 0\le {{x}_{nk}}\le 1,\ \ \forall n\in \mathcal{N},\forall k\in \mathcal{K} \nonumber
\end{IEEEeqnarray}
\par
Considering high complexity of the problem \eqref{eq11}, we introduce another variable ${{y}_{n}},\forall n\in \mathcal{N}$ to change it into a simple joint optimization problem.
\begin{equation}\label{eq12}
\begin{IEEEeqnarraybox}[][c]{cl}
  \underset{\mathbf{x},\mathbf{y}}{\mathop{ \max }}&\,\sum\limits_{n\in \mathcal{N}}{\sum\limits_{k\in \mathcal{K}}{{{x}_{nk}}{{s}_{nk}}\left\{ \log \left( {{R}_{nk}} \right)-\log \left( {{y}_{n}} \right) \right\}}} \\
 \text{    s}\text{.t}\text{.   }&\sum\limits_{n\in \mathcal{N}}{{{x}_{nk}}}=1,\ \ \forall k\in \mathcal{K} \\
 & \sum\limits_{k\in \mathcal{K}}{{{x}_{nk}}{{s}_{nk}}}={{y}_{n}},\ \ \ \forall n\in \mathcal{N} \\
 & {{y}_{n}}\le M\text{,}\ \ \forall n\in \mathcal{N} \\
 & 0\le {{x}_{nk}}\le 1,\ \ \forall n\in \mathcal{N},\forall k\in \mathcal{K}
\end{IEEEeqnarraybox}
\end{equation}
where $\mathbf{y}=\left\{ {{y}_{n}},n\in \mathcal{N} \right\}$ represents the number of consumed resources on BS $n$, i.e. load.
\section{ASSOCIATION ALGORITHM}
In this section, we give two types of algorithms to solve the problem \eqref{eq12}, which include a centralized algorithm by relaxing variables and a distributed algorithm by decoupling constraints.
\subsection{Max-Probability Association}
Before giving the max-probability association algorithm, we need to make the following definition.
\begin{Definition}
{\itshape If ${{x}_{nk}}$ is relaxed from the integer domain to region $\left[ 0,1 \right]$, the relaxed value is the probability that user $k$ is associated with BS $n$, i.e., association probability.}
\end{Definition}
\par
By relaxing association indicator variables, the combinatorial problem is converted into a convex optimization problem. Then, we can obtain a max-probability association algorithm that user is only associated with some BS having maximal association probability value obtained from solutions of the convex optimization problem, which can provide a pseudo-optimal solution that approaches a global optimal solution located at the boundary of the feasible region \cite{28}.
\par
For the optimal solution of the problem \eqref{eq12}, the corresponding Lagrange function can be expressed as
\begin{equation}\label{eq13}
\begin{IEEEeqnarraybox}[][c]{l}
  L\left( \mathbf{x},\mathbf{y},\mathbf{\gamma} ,\mathbf{\lambda} ,\mathbf{\mu} ,\mathbf{\nu}  \right)\\
  =\sum\limits_{n\in \mathcal{N}}{\sum\limits_{k\in \mathcal{K}}{{{x}_{nk}}{{s}_{nk}}\left( \log \left( {{R}_{nk}} \right)-\log \left( {{y}_{n}} \right) \right)}} \\
 \ \ \ +\sum\limits_{k\in \mathcal{K}}{{{\gamma }_{k}}( 1-\sum\limits_{n\in \mathcal{N}}{{{x}_{nk}}} )}
+\sum\limits_{n\in \mathcal{N}}{\sum\limits_{k\in \mathcal{K}}{{{\nu }_{nk}}\left( 1-{{x}_{nk}} \right)}}\\
  \ \ \ +\sum\limits_{n\in \mathcal{N}}{{{\lambda }_{n}}( {{y}_{n}}-\sum\limits_{k\in \mathcal{K}}{{{x}_{nk}}{{s}_{nk}}} )}+\sum\limits_{n\in \mathcal{N}}{{{\mu }_{n}}\left( M-{{y}_{n}} \right)}
\end{IEEEeqnarraybox}
\end{equation}
where $\mathbf{\gamma} =\left\{ {{\gamma }_{k}},k\in \mathcal{K} \right\}$, $\mathbf{\lambda} =\left\{ {{\lambda }_{n}},n\in \mathcal{N} \right\}$, $\mathbf{\mu} =\left\{ {{\mu }_{n}},n\in \mathcal{N} \right\}$ and $\mathbf{\nu} =\left\{ {{\nu }_{nk}},n\in \mathcal{N},k\in \mathcal{K} \right\}$ are the Lagrange multiplier vectors. The objective function of the dual problem can be defined as
\begin{equation}\label{eq14}
\begin{IEEEeqnarraybox}[][c]{l}
LD\left( \gamma ,\lambda ,\mu ,\nu  \right)=\underset{\mathbf{x}\ge 0,\mathbf{y}>0}{\mathop{\max }}\,L\left( \mathbf{x},\mathbf{y},\gamma ,\lambda ,\mu ,\nu  \right)
\end{IEEEeqnarraybox}
\end{equation}
and a dual problem of the primal problem defined in \eqref{eq12} is
\begin{equation}\label{eq15}
\begin{IEEEeqnarraybox}[][c]{l}
\underset{\gamma ,\lambda ,\mu \ge 0,\nu \ge 0}{\mathop{\min }}\,LD\left( \gamma ,\lambda ,\mu ,\nu  \right)
\end{IEEEeqnarraybox}
\end{equation}
\par
Since the primal problem \eqref{eq12} is a convex optimization problem,  a strong duality exists \cite{29}. Thus, the optimal solutions for the primal and dual problem are equal. Therefore, it is feasible to solve \eqref{eq12} by using the dual problem \eqref{eq15}. The problem refers to \eqref{eq14} can be further simplified to
\begin{equation}\label{eq16}
\begin{IEEEeqnarraybox}[][c]{l}
  LD\left( \gamma ,\lambda ,\mu ,\nu  \right) \\
 =\sum\limits_{n\in \mathcal{N}}{\underset{\mathbf{x}\ge 0,\mathbf{y}>0}{\mathop{\max }}\,\left\{ \sum\limits_{k\in \mathcal{K}}{{{x}_{nk}}{{s}_{nk}}\left( \log \left( {{R}_{nk}} \right)-\log \left( {{y}_{n}} \right) \right)} \right.} \\
 \ \ \ \left. +\left( {{\lambda }_{n}}-{{\mu }_{n}} \right){{\text{y}}_{n}}-\sum\limits_{k\in \mathcal{K}}{\left( {{\gamma }_{k}}+{{\lambda }_{n}}{{s}_{nk}}+{{\nu }_{nk}} \right){{x}_{nk}}} \right\}
\end{IEEEeqnarraybox}
\end{equation}
\par
Consequently, each BS can separately solve its own utility maximization problem, which is expressed as
\begin{equation}\label{eq17}
\begin{IEEEeqnarraybox}[][c]{l}
F\left( {{x}_{nk}},{{y}_{n}} \right) \\
 =\underset{\mathbf{x}\ge 0,\mathbf{y}>0}{\mathop{\max }}\,\left\{ \sum\limits_{k\in \mathcal{K}}{{{x}_{nk}}{{s}_{nk}}\left( \log \left( {{R}_{nk}} \right)-\log \left( {{y}_{n}} \right) \right)} \right. \\
 \ \ \ +\left( {{\lambda }_{n}}-{{\mu }_{n}} \right){{\text{y}}_{n}}\left. -\sum\limits_{k\in \mathcal{K}}{\left( {{\gamma }_{k}}+{{\lambda }_{n}}{{s}_{nk}}+{{\nu }_{nk}} \right){{x}_{nk}}} \right\}
\end{IEEEeqnarraybox}
\end{equation}
\par
Given the values of these Lagrange multipliers, the association indicator and load of BS can be obtained by applying Karush-Kuhn-Tucker (KKT) conditions on \eqref{eq17} \cite{30}, and we have
\begin{equation}\label{eq18}
\begin{IEEEeqnarraybox}[][c]{l}
{{x}_{nk}}={{\left[ \frac{\left( {{\lambda }_{n}}-{{\mu }_{n}} \right){{y}_{n}}}{{{s}_{nk}}} \right]}^{+}}
\end{IEEEeqnarraybox}
\end{equation}
\begin{equation}\label{eq19}
\begin{IEEEeqnarraybox}[][c]{l}
{{y}_{n}}=\exp \left( \log \left( {{R}_{nk}} \right)-\frac{{{\gamma }_{k}}+{{\lambda }_{n}}{{s}_{nk}}+{{\nu }_{nk}}}{{{s}_{nk}}} \right)
\end{IEEEeqnarraybox}
\end{equation}
where $\exp \left( z \right)$ is an exponential function of the $z$ and the notion ${{\left[ z \right]}^{+}}$ represents a projection on the positive orthant, which is used to account for the case that $z\ge 0$.
The optimum values of the above Lagrange multipliers that provide the optimum load distribution can be calculated by solving the dual problem \eqref{eq15}. As the dual function is differentiable, the Lagrange multipliers ${{\gamma }_{k}}$, ${{\lambda }_{n}}$, ${{\mu }_{n}}$ and ${{\nu }_{nk}}$ can be obtained with a gradient descent, and given by
\begin{IEEEeqnarray}{ll}
\gamma _{k}^{t+1}&=\gamma _{k}^{t}-{{\xi }_{1}}\left( 1-\sum\nolimits_{n\in \mathcal{N}}{x_{nk}^{t+1}} \right)\label{eq20}\\
\lambda _{n}^{t+1}&=\lambda _{n}^{t}-{{\xi }_{2}}\left( y_{n}^{t+1}-\sum\nolimits_{k\in \mathcal{K}}{{{s}_{nk}}x_{nk}^{t+1}} \right)\label{eq21}\\
\mu _{n}^{t+1}&={{\left[ \mu _{n}^{t}-{{\xi }_{3}}\left( M-y_{n}^{t+1} \right) \right]}^{+}}\label{eq22}\\
\nu _{nk}^{t+1}&={{\left[ \nu _{nk}^{t}-{{\xi }_{4}}\left( 1-x_{nk}^{t+1} \right) \right]}^{+}}\label{eq23}
\end{IEEEeqnarray}
where ${{\xi }_{1}}$, ${{\xi }_{2}}$, ${{\xi }_{3}}$ and ${{\xi }_{4}}$ are sufficiently small fixed step size for updating ${{\gamma }_{k}}$, ${{\lambda }_{n}}$, ${{\mu }_{n}}$ and ${{\nu }_{nk}}$, respectively. There is a convergence guarantee for the optimum solution since the gradient of the problem \eqref{eq15} satisfies the Lipchitz continuity condition \cite{30}. As a result, the ${{x}_{nk}}$ in problem \eqref{eq18} can achieve an optimum solution with a convergence guarantee.
\par
As shown in algorithm \ref{alg1}, after getting the optimum association probability values, each user selects a BS having maximum association probability value ${{x}_{nk}}$. If users are associated with BSs by using all optimum association indicators instead of only selecting the maximum value, a fractional user association algorithm that associates one user to many BSs will be found. In reality, the fractional user association scheme is often inadvisable because of its high implementation complexity.
\begin{algorithm}
\caption{Max-probability Association}
\label{alg1}
\begin{algorithmic}[1]
\STATE Initialize step size $\xi $ and estimate ${{R}_{nk}},\forall n\in \mathcal{N},\forall k\in \mathcal{K}$
\IF{$t=0$}
\STATE Initialize association probability $x_{nk}^{0}$ and Lagrange multipliers including $\gamma _{k}^{0}$, $\lambda _{n}^{0}$, $\mu _{n}^{0}$ and $\nu _{nk}^{0}$.
\ELSE
\STATE Calculate $y_{n}^{t+1}$ and $x_{nk}^{t+1}$ by applying KKT conditions with \eqref{eq18}-\eqref{eq19} .
\IF{iteration reaches the convergence precision (condition) of ${{x}_{nk}}$  or the maximum iteration number}
\STATE Set the maximal value of optimal association probabilities ($x_{n:}^{t+1}$) of each BS($n$) to 1, and transmit it to the corresponding BS($n$).
\ELSE
\STATE Update Lagrange multipliers via \eqref{eq20}-\eqref{eq23} utilizing  information ${{x}_{nk}}$ and ${{y}_{n}}$.
\STATE $t\leftarrow t+1$
\ENDIF
\ENDIF
\end{algorithmic}
\end{algorithm}
\par
In order to solve the convex optimization problem \eqref{eq12} directly, global network information should be provided, which means a centralized controller is needed to carry out user association and coordination. Considering this case, we give a distributed association algorithm without coordination in the subsequent section.
\subsection{Distributed Association}
Since the centralized functionality that solves the convex optimization problem tends towards high computational complexity and low reliability, a low overhead and complexity distributed algorithm is desirable.
\par
In this section, we utilize the Lagrange dual decomposition method \cite{31} to develop a distributed algorithm, which is in accord with Ye's \cite{16}. Instead of directly solving the original convex optimization problem \eqref{eq12}, users and BSs can separately solve two sub-problems which the dual problem is decoupled into.
\par
The primal problem \eqref{eq12} can be rewritten in an equivalent form as follows:
\begin{equation}\label{eq24}
\begin{IEEEeqnarraybox}[][c]{ll}
   \underset{\mathbf{x},\mathbf{y}}{\mathop{ \max }}&\,\sum\limits_{n\in \mathcal{N}}{\sum\limits_{k\in \mathcal{K}}{{{x}_{nk}}{{s}_{nk}}\log \left( {{R}_{nk}} \right)-}}\sum\limits_{n\in \mathcal{N}}{{{y}_{n}}\log \left( {{y}_{n}} \right)} \\
  \text{    s}\text{.t}\text{.   }&\sum\limits_{n\in \mathcal{N}}{{{x}_{nk}}}=1,\ \ \forall k\in \mathcal{K} \\
 & \sum\limits_{k\in \mathcal{K}}{{{x}_{nk}}{{s}_{nk}}}={{y}_{n}},\ \ \forall n\in \mathcal{N} \\
 & {{y}_{n}}\le M,\ \ \forall n\in \mathcal{N}\\
 & 0\le {{x}_{nk}}\le 1,{{y}_{n}}>0,\ \ \forall n\in \mathcal{N},\forall k\in \mathcal{K}
\end{IEEEeqnarraybox}
\end{equation}
\par
Obviously, there is no coupling constraint but $\sum\nolimits_{k\in \mathcal{K}}{{{x}_{nk}}}={{y}_{n}}$ in problem \eqref{eq24}, which motivates us to  convert the above problem through introducing a Lagrange multiplier $\mu $ to relax this coupled constraint. Then the dual problem can be expressed as
\begin{equation}\label{eq25}
\begin{IEEEeqnarraybox}[][c]{l}
\mathbf{G: \ }\underset{\mu }{\mathop{\min }}\,\ G\left( \mu  \right)=H\left( \mu  \right)+I\left( \mu  \right)
\end{IEEEeqnarraybox}
\end{equation}
where
\begin{equation}\label{eq26}
H\left( \mu  \right)=\left\{ \,
\begin{IEEEeqnarraybox}[][c]{ll}
\IEEEstrut
 \underset{\mathbf{x}}{\mathop{ \max }}&\,\ \sum\limits_{n\in \mathcal{N}}{\sum\limits_{k\in \mathcal{K}}{{{x}_{nk}}{{s}_{nk}}\left\{ \log \left( {{R}_{nk}} \right)-{{\mu }_{n}} \right\}}} \\
 \text{s}\text{.t}\text{.  }&\sum\limits_{n\in \mathcal{N}}{{{x}_{nk}}}=1,\ \ \forall k\in \mathcal{K} \\
 &{{x}_{nk}}\in \left\{ 0,1 \right\},\ \ \forall n\in \mathcal{N},\forall k\in \mathcal{K}
\IEEEstrut
\end{IEEEeqnarraybox}
\right.
\end{equation}
\begin{equation}\label{eq27}
\begin{IEEEeqnarraybox}[][c]{l}
I\left( \mu  \right)=\underset{{{y}_{n}}\le M}{\mathop{ \max }}\,\ \sum\limits_{n\in \mathcal{N}}{{{y}_{n}}\left\{ {{\mu }_{n}}-\log \left( {{y}_{n}} \right) \right\}}
\end{IEEEeqnarraybox}
\end{equation}
\par
There is a strong duality if the above two sub-problems (inner problems) have the same optimal solution. Being a sufficient condition for holding strong duality in a convex optimization problem, Slater condition becomes feasibility if it only includes linear equalities and inequalities \cite{30}. Evidently, the primal problem \eqref{eq24} meets the qualification due to its constraints are all linear. Thus, the problem can be equivalently converted into a dual problem \eqref{eq25}. Denoting ${{x}_{nk}}\left( \mu  \right)$ and ${{y}_{n}}\left( \mu  \right)$ as the maximizers for sub-problem \eqref{eq26} and \eqref{eq27}, respectively. A dual optimal ${{\mu }^{*}}$ that results in the primal optimal values ${{x}_{nk}}\left( {{\mu }^{*}} \right)$ and ${{y}_{n}}\left( {{\mu }^{*}} \right)$ can be obtained. Therefore, we can easily find the primal optimal solution for a given the dual optimal ${{\mu }^{*}}$  by solving separately its two sub-problems without coordination among the users and BSs.
\par
Now we solve the outer problem \eqref{eq25} using a gradient projection method \cite{32}, where the Lagrange multiplier $\mu $ is updated in the direction of the negative gradient, i.e., $-\nabla G\left( \mu  \right)$. Evaluating the gradient of dual objective function relates to solving the minimization problem \eqref{eq25}, which has been equivalently converted into two sub-problems that can be dealt in a distributed manner.
\par
According to the direct observation, the problem \eqref{eq26} can be further simplified to
\begin{equation}\label{eq28}
\begin{IEEEeqnarraybox}[][c]{l}
\underset{n}{\mathop{\max }}\,\left\{ {{s}_{nk}}\left( \log \left( {{R}_{nk}} \right)-{{\mu }_{n}} \right) \right\}
\end{IEEEeqnarraybox}
\end{equation}
which means each user chooses one BS to maximize the utility ${{s}_{nk}}\left( \log \left( {{R}_{nk}} \right)-{{\mu }_{n}} \right)$. It is indeed an algorithm at user's side, which is illustrated as algorithm \ref{alg2}.

\begin{algorithm}
\caption{at User Terminal $k$}
\label{alg2}
\begin{algorithmic}[1]
\IF{$t=0$}
\STATE Estimate ${{R}_{nk}},\forall n\in \mathcal{N}$ using pilot signals from all BSs.
\ELSE
\STATE User $k$ receives $\mu _{n}^{t},\forall n\in \mathcal{N}$ broadcasted by all BSs and is associated with BS ${{n}^{*}}$ which satisfies the following formula:
$${{n}^{*}}=\arg \underset{n}{\mathop{\max }}\,\left\{ {{s}_{nk}}\left( \log \left( {{R}_{nk}} \right)-\mu _{n}^{t} \right) \right\}$$
If there are many optimum associations at the same time, user can choose any one of them.
\STATE Feedback association information ${{x}_{{{n}^{*}}k}}=1$ to BS ${{n}^{*}}$.
\ENDIF
\end{algorithmic}
\end{algorithm}
\par
For the problem \eqref{eq27}, the optimum load ${{y}_{n}}$ can be calculated for each BS by applying KKT condition, we have
\begin{equation}\label{eq29}
\begin{IEEEeqnarraybox}[][c]{l}
{{y}_{n}}=\min \left\{ {{e}^{{{\mu }_{n}}-1}},M \right\}
\end{IEEEeqnarraybox}
\end{equation}
\par
After getting the optimum load ${{y}_{n}}$ at time slot $t$, the Lagrange multiplier ${{\mu }_{n}}$ can be updated by each BS with
\begin{equation}\label{eq30}
\begin{IEEEeqnarraybox}[][c]{l}
\mu _{n}^{t+1}=\mu _{n}^{t}-\xi \left( y_{n}^{t}-\sum\limits_{k\in \mathcal{K}}{{{s}_{nk}}x_{nk}^{t}} \right)
\end{IEEEeqnarraybox}
\end{equation}
where $\xi$ is a  sufficiently small fixed step size for updating the ${{\mu }_{n}}$. The detailed procedure refers to calculating the optimum load ${{y}_{n}}$ and updating the Lagrange multiplier ${{\mu }_{n}}$, which is provided in algorithm \ref{alg3}.

\begin{algorithm}
\caption{at Base Station $n$}
\label{alg3}
\begin{algorithmic}[1]
\IF{$t=0$}
\STATE Initialize step size $\xi $ and $\mu _{n}^{t}$.
\ELSE
\STATE Receive  association information $x_{nk}^{t}=1,\forall k\in \mathcal{K}$ and calculate $y_{n}^{t}$ by applying KKT condition on problem \eqref{eq27}.
$$y_{n}^{t}=\min \left\{ \exp \left( \mu _{n}^{t}-1 \right),M \right\}$$
\STATE Update the Lagrange multiplier $\mu _{n}^{t+1}$ using association information.
$$\mu _{n}^{t+1}=\mu _{n}^{t}-\xi \left( y_{n}^{t}-\sum\limits_{k\in \mathcal{K}}{{{s}_{nk}}x_{nk}^{t}} \right)$$
\STATE Broadcast the new $\mu _{n}^{t+1}$ value to all users.
\ENDIF
\end{algorithmic}
\end{algorithm}
\par
Through a further observation, we find factors of the formula \eqref{eq30} have some interesting meanings. The multiplier $\mu $ can be regarded as a message between users and BSs, and interpreted as the service cost of BSs decided by the load distribution. It tradeoffs  supply and demand if $\sum\nolimits_{k\in \mathcal{K}}{{{s}_{nk}}{{x}_{nk}}}$ is deemed as the serving demand for BS $n$ and ${{y}_{n}}$ is considered as the service BS $n$ can provide. In fact, the formula \eqref{eq30} meets the law of supply and demand, which means that the service cost will go up if the demand $\sum\nolimits_{k\in \mathcal{K}}{{{s}_{nk}}{{x}_{nk}}}$ for BS $n$ exceeds the supply ${{y}_{n}}$ and vice versa. Therefore, some BS will increase its service price so that fewer users are associated with itself if it is over-loaded, while other under-loaded ones will decrease the service price to attract more users.
\par
It is very easy to find that adjusting $\mu $ with the formula \eqref{eq30} is made completely distributed among BSs and only relies on local information. For each iteration, the distributed method shows that algorithms at the user's and BS's side have computation complexity of $\mathcal{O}\left( N \right)$ and $\mathcal{O}\left( K \right)$ respectively. Compared with the distributed algorithm, the centralized algorithm appears to be more complicated, which has computation complexity of $\mathcal{O}\left( NK \right)$ for each iteration. With regard to the exchanged information involved in each iteration, each BS broadcasts its service cost $\mu $ that contains relatively little information to all users, and each user reports its service request to only one BS which it expects to connect to. At each iteration, the distributed algorithm owns the amount of exchanged information of $K+N$, while the amount in centralized method is proportional to $K\times N$. Moreover, due to the high convergence speed of the distributed algorithm, the iteration number is small. Consequently,  the distributed algorithm may be more practicable for some cases, especially for large scale problems, even if there exits more exchanged information. It is effective as long as the distributed algorithm can converge to a near-optimal solution in a given association period. After interactively carrying out the algorithms at the user's and BS's side, the distributed method can finally provide a near-optimal solution, which will be proved in the following theorem.
\begin{Theorem}
{\itshape If ${{G}^{*}}>-\infty $ where ${{G}^{*}}$ denotes the optimal value of the problem \eqref{eq25}, then}
\end{Theorem}
\begin{equation}\label{eq31}
\begin{IEEEeqnarraybox}[][c]{l}
\underset{t}{\mathop{\inf }}\,G\left( {{\mu }^{t}} \right)\le {{G}^{*}}+\varphi
\end{IEEEeqnarraybox}
\end{equation}
\itshape{Proof:} \upshape The derivative of the function ${G}$ is calculated by
\begin{equation}\label{eq32}
\begin{IEEEeqnarraybox}[][c]{l}
\frac{\partial G}{\partial {{\mu }_{n}}}\left( \mu  \right)={{y}_{n}}\left( \mu  \right)-\sum\limits_{k\in \mathcal{K}}{{{s}_{nk}}{{x}_{nk}}\left( \mu  \right)}
\end{IEEEeqnarraybox}
\end{equation}
\par
Obviously, the function $\partial G$ is bounded because ${{y}_{n}}$ and $\sum\nolimits_{k\in \mathcal{K}}{{{s}_{nk}}{{x}_{nk}}}$ are bounded, where ${{y}_{n}}\le M$. Therefore, our problem meets the necessary conditions of proposition 6.3.6 in \cite{31}, and the theorem can be proved by applying this proposition.
\section{PERFORMANCE EVALUATION}
\subsection{Simulation Setup}
Without loss of generality, we only consider a two-tier HetNet with transmit power $\left\{ {{P}_{1}},{{P}_{2}} \right\}=\left\{ 46,20 \right\}$ dBm, which includes macro and pico BSs. The location of macro BSs is modeled to be fixed and forms a conventional cellular structure, while pico BSs are uniformly and independently distributed in macrocells. The users are scattered into macrocells in the same way as pico BSs. In modeling the propagation environment, we adopt the following general path loss model \cite{33}
\begin{equation}\label{eq33}
\begin{IEEEeqnarraybox}[][c]{l}
PL\left( d \right)=20\log \left( \frac{4\pi d}{\lambda } \right)+10n\log \left( \frac{d}{{{d}_{0}}} \right)
\end{IEEEeqnarraybox}
\end{equation}
where $d$ is the distance (m) between the transmitter and receiver; $\lambda $ denotes the wavelength of radiation (m); ${{d}_{0}}$ represents a reference distance at which or closer to the path loss inherits the characteristics of free-space loss in equation (1.2) \cite{33}, which equals 50 m and 1m for macro and pico system, respectively; $n$ is the path loss exponent and respectively set to be 3 and 3.5 for macrocell and picocell.
\par
As for another large-scale fading, i.e., shadowing, we assume lognormal shadowing with standard deviation 8dB and 10dB for macro and pico BSs respectively. Meanwhile, we consider that there is noise power of 174dBm/Hz, system bandwidth of 20MHz and carrier frequency of 2GHz.
\par
In the following simulation sections, we will compare user-based and resource-based methods in two cases including users having different practical rates and having identical practical rates. Note that, if we adopt an association scheme without regarding the restriction of resources, there may be more associated users than actual served users and some associated users who can't be served by the corresponding BS. To give a good explanation for  the phenomenon by numerical simulation, we investigate the performance which concerns the call blocking probability and load balancing index obtained by adopting above two kinds of association algorithms. For this purpose, we assume that each BS schedules served users from its associated user queue according to maximal practical rate first (MPRF) that selects users  in descending order of practical rates and maximal achievable rate first (MARF) that selects users in descending order of achievable rates under the restriction of resources. In fact, whatever scheduling strategy is adopted, QoS-aware (resource-based) algorithms are often superior to user-based ones that ignore QoS requirements in terms of the call blocking probability and load balancing index. There are some reasons for this. First, users are associated with some BS according to their rate requirements under the actual restriction of resources for the former, while the latter doesn't consider any practical condition. Secondly, each BS selects served users under this restriction. In other words, the user-based algorithms achieve an illusive balance.
\subsection{Call Blocking Probability}
The call blocking probability mentioned in this paper can be expressed as
\begin{equation}\label{eq34}
\begin{IEEEeqnarraybox}[][c]{l}
Pr=1-\frac{u}{\left| \mathcal{K} \right|}
\end{IEEEeqnarraybox}
\end{equation}
where $u$ denotes the number of users scheduled from associated user queue; $\left| \mathcal{K} \right|$ represents total number of users scattered in HetNets.
\par
For the user-based method, the call blocking probability arises from the shortage of resources for some BSs, which means that some associated users may not communicate with the corresponding BS. However, for the resource-based method, some users can't be associated with some BS due to the insufficient resources of BSs, which results in another kind of call blocking probability. Comparing these two methods, we can easily find that their common ground lies in the cause of call blocking probability and difference comes from blocked users.
\par
\begin{figure}[!t]
\centering
\centerline{\includegraphics{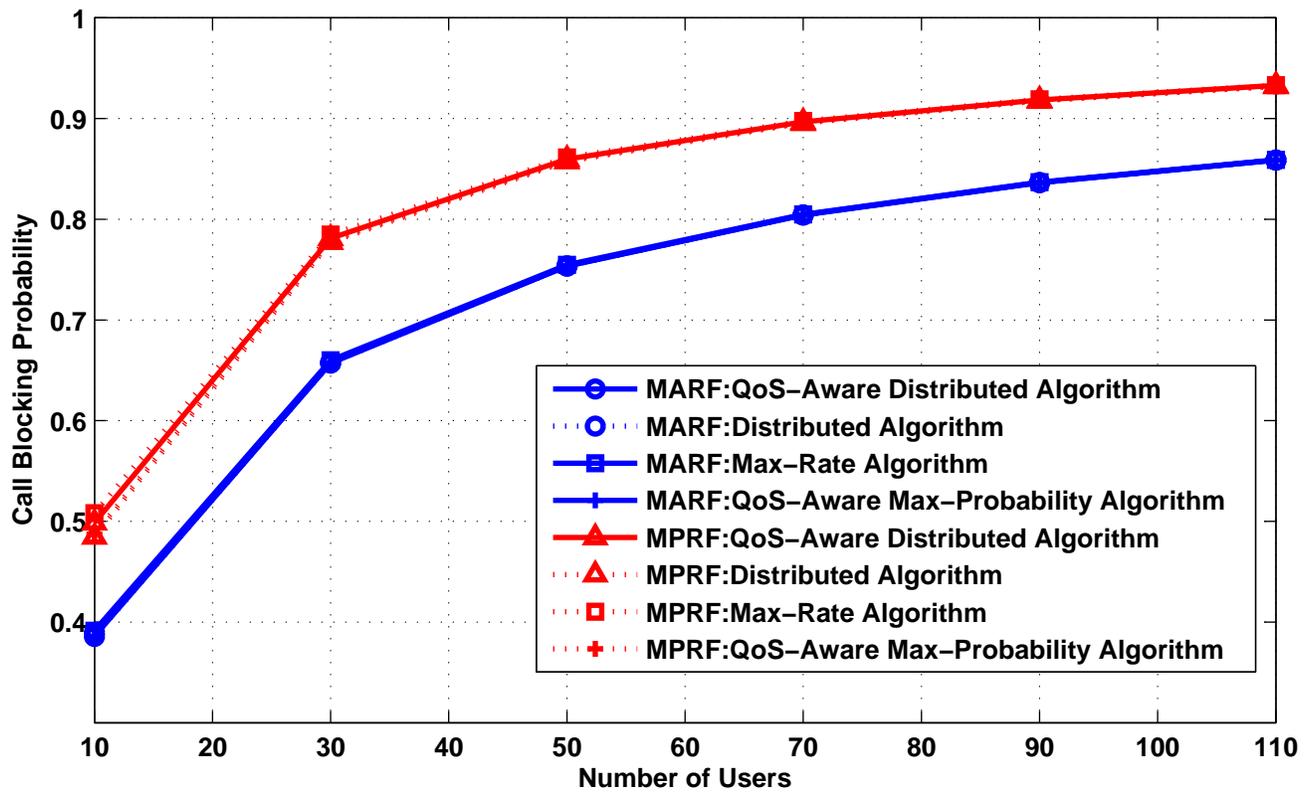}}
\caption{Call blocking probability for users having same rate requirements.}
\label{fig2}
\end{figure}

Fig.\ref{fig2} shows that four kinds of algorithms have almost the same performance when there are users having identical rate requirements (1 Mbps). It is apparent that two types of distributed association algorithms are actually equivalent since users having same actual rate need an equal amount of resources. Meanwhile, since most users' achievable rates often tend to be far smaller than the above actual rates, more resources should be supplied for supporting their practical operations. However, the amount of resources of each BS is limited, which results in very few users can be severed and almost the same call blocking rate between max-rate association and distributed association. Note that, the call blocking probability of MARF is distinctly less than MPRF, which increases with increasing number of users scattered in each macrocell because of the restriction of resources. Since a few achievable rates may be far bigger than actual rates when users are very close to high-power BSs, some users having higher achievable rates, who require fewer resources for meeting actual rate requirements, may be associated. Moreover, we also find that the cellular system accepts very few users due to high actual rate requirements and most low achievable rates. Therefore, more users having higher achievable rate (i.e., fewer resource requirement) are admitted if we select users as customers by MAPF but MPRF under the constraint of resources, which means a lower call blocking probability.
\par
\begin{figure}[!t]
\centerline{\includegraphics{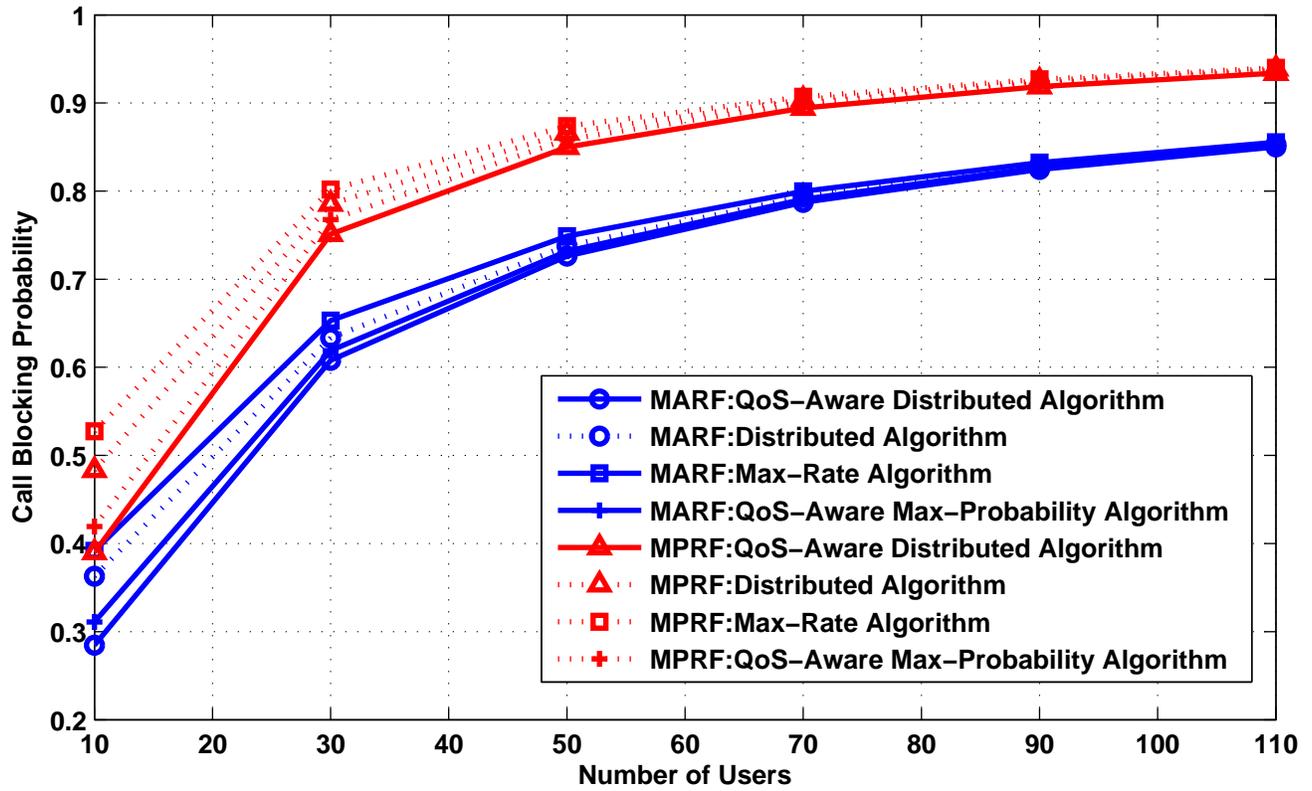}}
\caption{Call blocking probability for users having distinct rate requirements.}
\label{fig3}
\end{figure}

Fig.\ref{fig3} plots the call blocking probability for users having distinct rate requirements (random value in range (0, 2] Mbps). In low density of users deployed in each macrocell, the resource-based  algorithms including QoS-aware distributed algorithm and max-probability algorithm are vastly superior to the user-based one, and their superiorities are becoming smaller with increasing density of users. As illustrated in Fig.\ref{fig3}, the max-rate association owns the worst performance in all algorithms, and the result of the max-probability association is slightly inferior to the QoS-aware distributed algorithm.
\subsection{Load Balancing Index}
To measure the status of load balancing of system, we bring the following Jain¡¯s fairness index
\begin{equation}\label{eq36}
\begin{IEEEeqnarraybox}[][c]{l}
\eta =\frac{{{\left( \sum\limits_{n\in \mathcal{N}}{{{\rho }_{n}}} \right)}^{2}}}{\left| \mathcal{N} \right|\sum\limits_{n\in \mathcal{N}}{\rho _{n}^{2}}}
\end{IEEEeqnarraybox}
\end{equation}
where $\sum\nolimits_{k\in \mathcal{K}}{{{x}_{nk}}{{s}_{nk}}}={{\rho }_{n}}$ represents the amount of consumed resources; $\left| \mathcal{N} \right|$ is the number of given cells in the network. A larger $\eta $  taken value from the interval $\left[ \frac{1}{\left| \mathcal{N} \right|},1 \right]$, which means a more balanced load distribution among the given cells.
\par
\begin{figure}[!t]
\centering
\centerline{\includegraphics{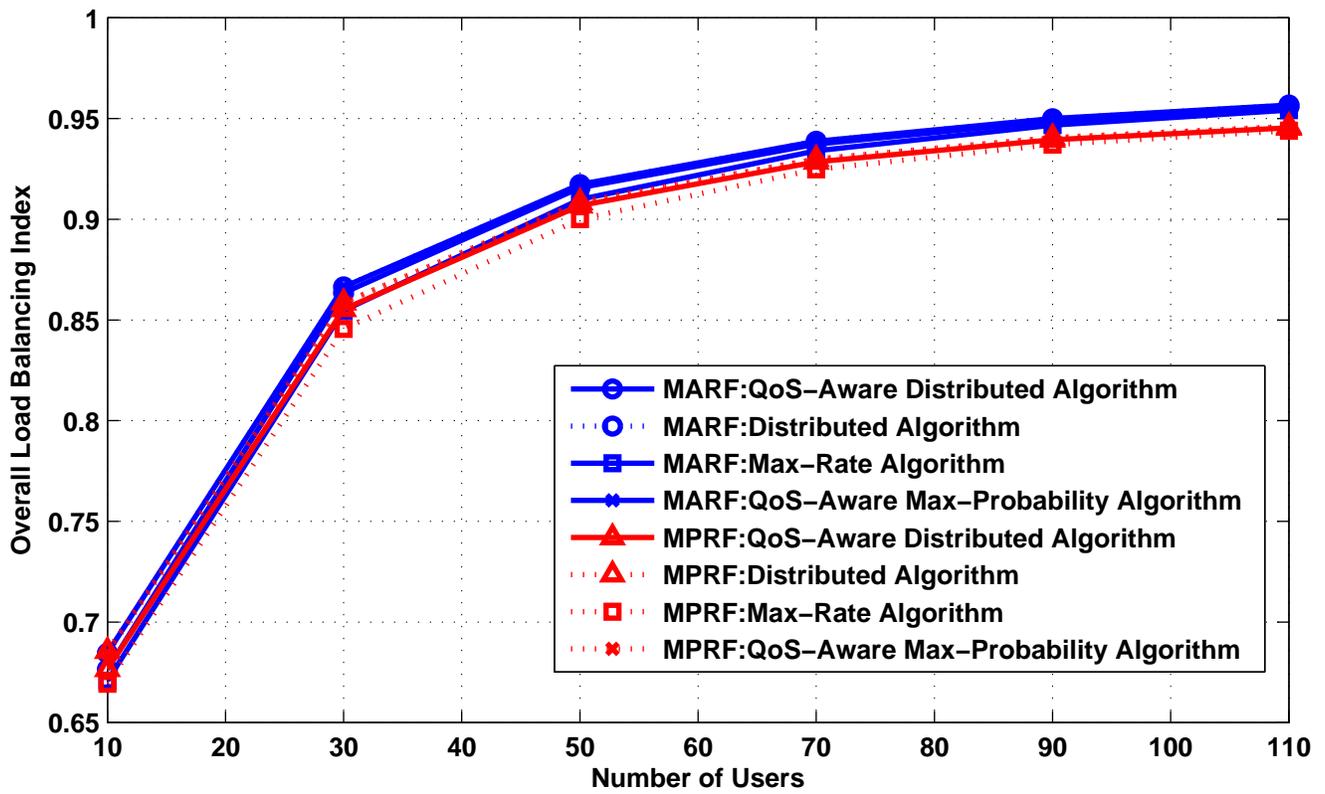}}
\caption{Overall load balancing index for users having same  rate requirements.}
\label{fig4}
\end{figure}

Fig.\ref{fig4} shows that four types of association algorithms are still maintaining almost the performance in terms of load balancing level of the entire network. Since common high practical rate requirement and most low achievable rates mean that more resources should be provided, very few users can be selected as customers under limited resources. Hence, the number of served users has a negligible effect on distributed algorithms. Note that, the load balancing index obtained by MARF is slightly higher than by MPRF, which is different from call blocking probability as far as curve relations are concerned. The main reason for this is that the load balancing index calculated by actual resources consumed by scheduled users and MARF selects users according to ascending requirements of resources. Obviously, the scheduling strategy (MARF) which utilizes ascending requirements of resources owns better performance than MPRF which tends to be random.
\par
\begin{figure}[!t]
\centering
\centerline{\includegraphics{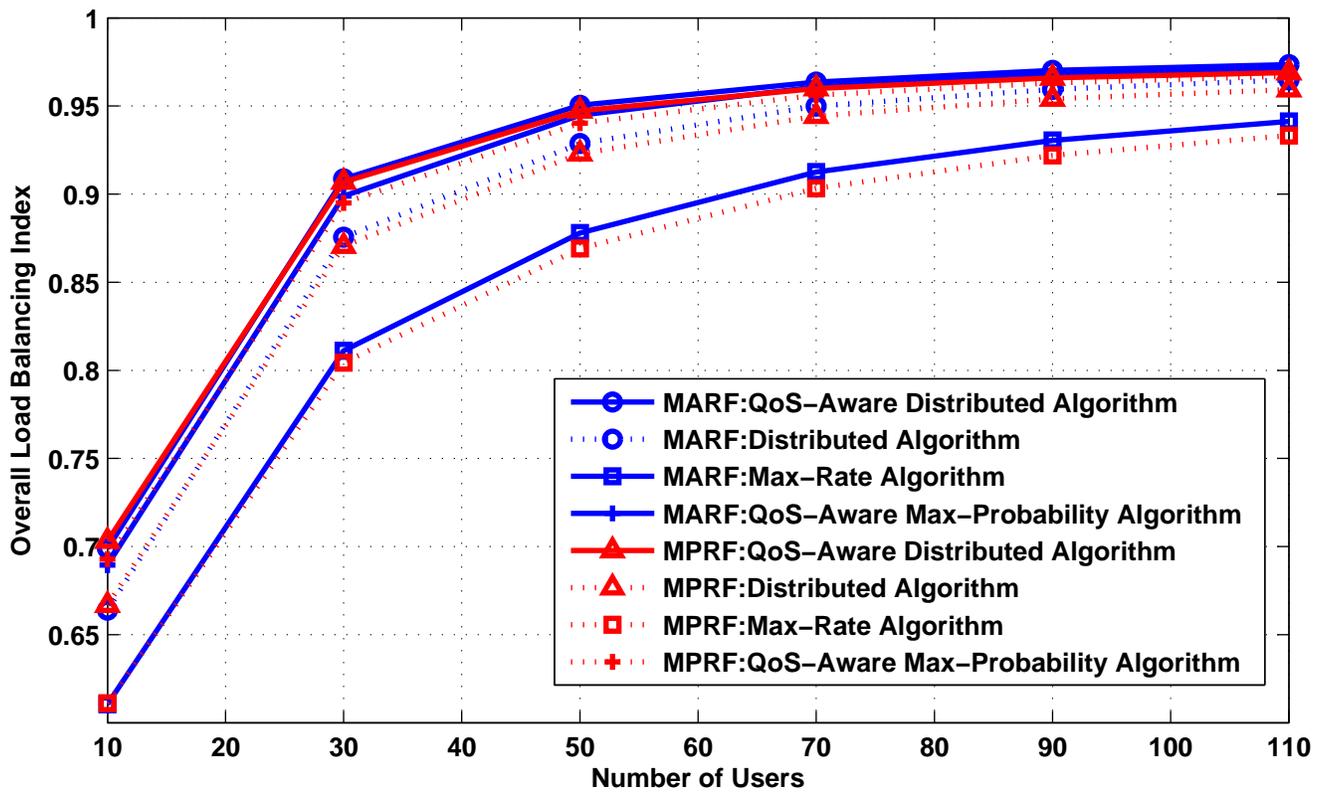}}
\caption{Overall load balancing index for users having distinct rate requirements.}
\label{fig5}
\end{figure}

As shown in the Fig.\ref{fig5}, resource-based algorithms including max-probability and QoS-aware distributed algorithm can provide  higher overall load balancing levels than others. Moreover, the max-rate association achieves the worst result in all algorithms, the performance of max-probability is slightly poorer than the corresponding distributed algorithm, and the scheduling scheme MARF brings a more favorable result than MPRF due to the fact mentioned in section A.
\par
\begin{figure}[!t]
\centering
\centerline{\includegraphics{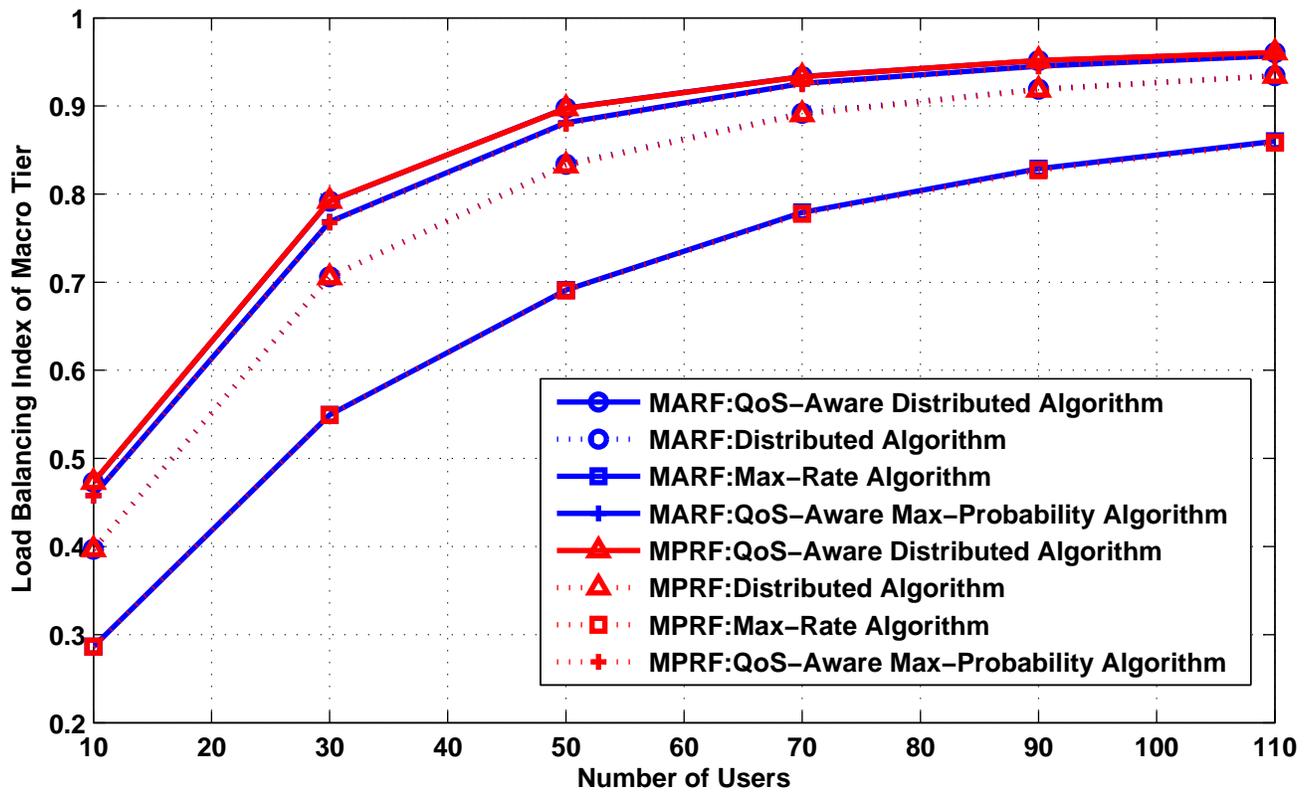}}
\caption{Load balancing index of macro tier for users having distinct rate requirements.}
\label{fig6}
\end{figure}

Fig.\ref{fig6} gives the load balancing index of the macro tier for users having different rate requirements. There are some same relations with overall load balancing index except for the relation between MARF and MPRF on four kinds of algorithms. As illustrated in Fig.6, the scheme MARF has almost same performance with MPRF, which is different from the case plotted by Fig.\ref{fig5}. Clearly, the overall load balancing index refers to all cells including macrocells and picocells (small cells), but the load balancing index of the macro tier only concentrates on all macrocells. As the lower transmit power and suffered severer interference of small cells  mean that more resources should be provided for supporting user's communication, fewer users can be severed by small cells. Moreover, the difference among interferences received by small cells may be very striking since we deploy small BSs in a random way. Consequently, small cells greatly affect the overall load balancing level of the system. However, the load balancing index of the macro tier is unaffected by small cells, meanwhile macrocells are on equal footing. Therefore, MARF can achieve the almost same result with MPRF in terms of the load level of the macro tier.
\subsection{Convergence}
\begin{figure}[!t]
\centering
\centerline{\includegraphics{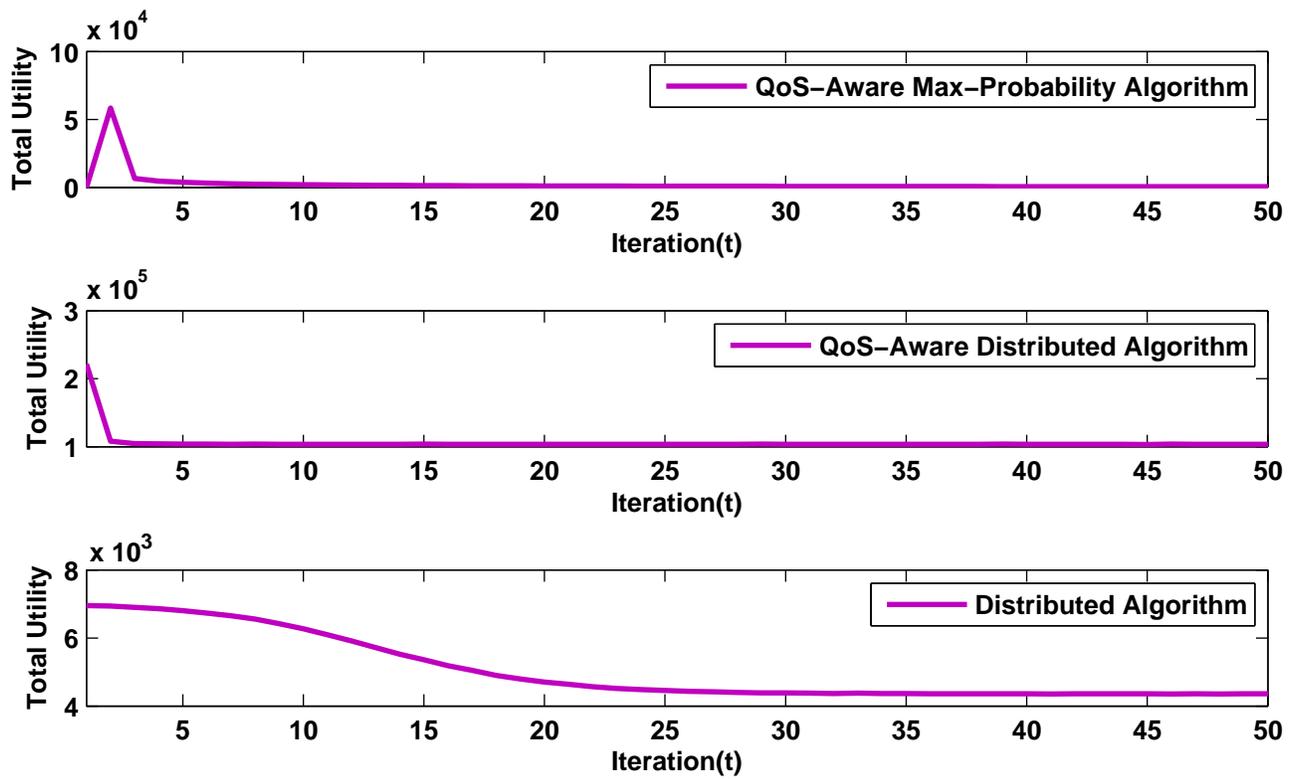}}
\caption{The convergent total (network-wide) utility of three algorithms.}
\label{fig7}
\end{figure}

Fig.\ref{fig7} compares the convergence of three algorithms, where parameter $t$ represents $t\text{-th}$ iteration. Note that two kinds of algorithms have different orders of magnitude on optimal total utilities due to their disparate objective functions. It can be seen that QoS-aware association algorithms own faster convergence rates than the algorithm without considering rate requirements. It is noteworthy that the number of iterations of two kinds of distributed algorithms actually represents the times of  information exchange between user and BS. The distributed algorithm we advocated can converge in just three iterations, which is far less than the distributed algorithm proposed by Ye. Therefore, our scheme can be well applied in reality.
\section{CONCLUSION}
In this paper, the load balancing problem for HetNets has been investigated in terms of services with QoS requirements. We first formulate the problem to be network-wide weighted utility of load efficiency maximization problem under constraint of resources. Then we give a low complexity distributed algorithm that is proved to converge to a near-optimal solution using dual decomposition and another near-optimal centralized association algorithm (i.e., max-probability association). After that the performance variance is looked into by utilizing two kinds of scheduling strategies according to different density of users scattered into each macrocell. The numerical results show that, the association algorithms we proposed have obvious superiorities over the user-based algorithm in terms of the call blocking probability, the load balancing index and the convergence rate. Future work can include developing dynamic association algorithms, considering the uplink scenario with power control, and introducing interference management techniques.
\bibliographystyle{IEEEtran}
\bibliography{reference}
%
\newpage

%
%






\end{document}